\documentclass{ws-mplb}
\usepackage{color}
\usepackage{graphicx}
\begin{document}

\markboth{Mauro M. Doria, Alfredo A. Vargas-Paredes, Jos\'e A.
Helay\"el Neto, }{The principle of local rotational invariance and
the coexistence of magnetism, charge and superconductivity}

%
\catchline{}{}{}{}{}
%

\title{THE PRINCIPLE OF LOCAL ROTATIONAL INVARIANCE AND THE COEXISTENCE OF MAGNETISM, CHARGE AND SUPERCONDUCTIVITY}
\author{\footnotesize Mauro M. Doria\footnote{ Av. Pedro Calmon, n° 550, RJ, Brazil}}

\address{Departamento de F\'{\i}sica dos S\'{o}lidos, Universidade
Federal do Rio de Janeiro, Av. Pedro Calmon, n° 550,\\
Rio de Janeiro 21941-972/RJ,
Brazil\\
\ mmd@if.ufrj.br}

\author{Alfredo A. Vargas-Paredes\footnote{Rua Dr. Xavier Sigaud, 150 - Urca, RJ, Brazil}}

\address{Laborat\'orio de F\'isica Experimental de Altas Energias, Centro Brasileiro de Pesquisas
F\'{\i}sicas, Rua Dr. Xavier Sigaud, 150 - Urca\\
Rio de Janeiro 22290-180/RJ, Brazil\\
\ alfredov@cbpf.br}

\author{\footnotesize Jos\'e A. Helay\"el-Neto\footnotemark[2]}

\address{Laborat\'orio de F\'isica Experimental de Altas Energias, Centro Brasileiro de Pesquisas
F\'{\i}sicas, Rua Dr. Xavier Sigaud, 150 - Urca\\
Rio de Janeiro 22290-180/RJ, Brazil\\
\ helayel@cbpf.br}

\maketitle

\begin{history}
\received{(Day Month Year)}
\revised{(Day Month Year)}
\end{history}

\begin{abstract}
We propose a macroscopic description of the superconducting state in
presence of an applied external magnetic field in terms of first
order differential equations. They describe a corrugated
two-component order parameter intertwined with a spin-charged
background, caused by spin correlations and charged dislocations.
The first order differential equations are a consequence of a
Weitzenb\"ock-Liechnorowitz identity which renders a
SU$_L$(2)$\otimes$ U$_L$(1) invariant ground state, based on (L)
local rotational and electromagnetic gauge symmetry. The proposal is
based on a long ago developed formalism by \'Elie Cartan to
investigate curved spaces, viewed as a collection of small Euclidean
granules that are translated and rotated with respect to each other.
\'Elie Cartan's formalism unveils the principle of local rotational
invariance as a gauge symmetry because the global $SU(2)$ invariance
of the order parameter is turned into a local invariance by the
interlacement of spin and charge to pairing.
\end{abstract}


\section{Introduction}
\begin{figure}[h]
\centering
\includegraphics[width=1.00\linewidth]{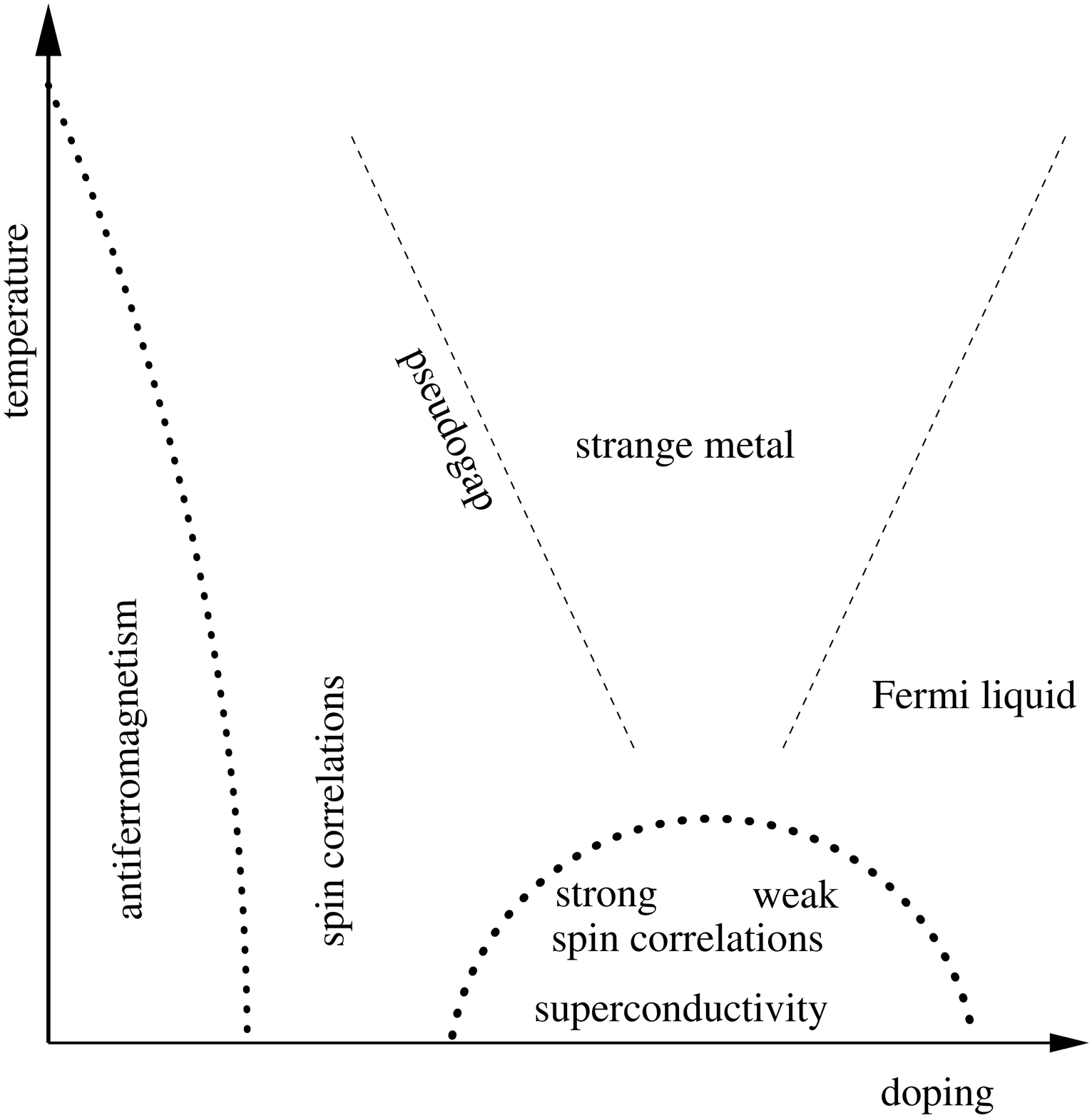}
\caption{A schematic phase diagram is shown here for the temperature
versus the concentration of charge carriers (electrons or holes),
called doping. For low doping the state is of an antiferromagnetic
insulator, but carriers become mobile by increasing doping, leading
to the pseudogap, the strange metal and the Fermi liquid domains,
which are displayed in the phase diagram. Spin correlations are
stronger near the antiferromagnetic insulator phase and are also
observed within the superconducting dome, although they become weak
upon doping.}\label{fig1}
\end{figure}

The present experimental evidence indicates a far reaching
complexity of the superconducting state in the new compounds not
found before in metals. Different electronic orders seem to coexist
inside the compound. It is even conceivable that a single set of
electrons participate in different orders, or more likely, that nano
separated neighbor electrons belong to different orders. The
evidence to the coexistence of different orders has been
accumulating since the early days of high-temperature ceramic
superconductors when physicists stomped into the
pseudogap~\cite{allaoul09}, a gap that sets in at a temperature much
above the critical temperature. Recent evidence shows that electrons
in the pseudogap phase are not paired up, but organized in a new
order that persists when the compound becomes
superconductor~\cite{he2011}. Another remarkable feature of
superconductivity in the new compounds is its
magnetism~\cite{lake02}. Magnetism was once thought to be
detrimental to the superconducting electron pairs, now is
acknowledged to coexist and possibly contribute to pair stability.
Nuclear magnetic resonance studies have revealed that spin
correlations coexist with superconductivity in the superconducting
cuprates varying from long-range to short-range correlations
according to doping, as depicted in Fig.\ref{fig1}. Therefore the
evidence is that superconductivity is intertwined with spin and
charge degrees of freedom~\cite{berg09} forming a highly correlated
electronic system. The onset of different types of inhomogeneous
states with broken rotational and translational symmetries, such as
striped, nematic and smectic phases~\cite{berg09,vojta09,daou10} can
be understood as a natural consequence of such coexistence. A
general theoretical framework able to deal with this plethora of
phenomena is still missing. Nevertheless the coexistence of spin,
charge, and pairing
orders~\cite{ando02,kanoda03,sanna10,tranquada11,koutroulakis10},
the evidence of multigap
superconductivity~\cite{hufner08,prozorov11,matano08,gurevich07},
and the layered structure seem to be common features of the new
compounds.

In this paper we consider the principle of local rotation as a way
to treat all these common features and describe the onset of
heterogenous states. The simplest possible framework to describe the
superconducting state is through the quantum macroscopic approach,
whose goal is to describe the superconducting state by means of an
order parameter (OP). Within this framework we study this principle
of local rotational invariance taking that the non-superconducting
degrees of freedom are spin correlations and charged dislocations
interlaced with the superconducting OP to form a heterogeneous
state. The superconducting OP feels the presence of the spin
correlations and charged dislocations to become spatially
corrugated. Reversely the corrugated OP act in these
non-superconducting degrees of freedom, but this is not treated
here, and for this reason we refer to the presence of these spin
correlations and charged dislocations as {\it the spin-charged
background}.

The quantum macroscopic approach is very successful to describe
superconducting phenomena and yet it was formulated without
knowledge that electrons pairs form the ground state of the
superconducting state, which is a key ingredient of the macroscopic
theory. In 1950, Vitalii Ginzburg and Lev Landau published their
phenomenological theory of superconductivity by including the
principle of gauge invariance into the general theory of the second
order phase transitions proposed earlier by Landau in 1937. For this
purpose, the OP was set complex in order to have minimal coupling to
the magnetic field, a puzzling assumption later proven to be
correct. The Meissner effect was explained on this basis, and so, it
can be regarded as a natural consequence of the gauge invariance of
the Ginzburg-Landau theory. Similarly, the present formalism
proposes a new gauge symmetry to explain the observed inhomogeneity
of the superconducting state, that is, to describe the spatial
corrugations of the OP in presence of the spin-charged background.
This new gauge invariance is the principle of local rotational
invariance to be described in this paper.

The ultimate goal of the quantum macroscopic approach is to
determine the OP, and in this way to describe the superconducting
state. Usually this is achieved by firstly proposing a free energy
expansion in powers of the OP, such as in the Ginzburg-Landau
theory. However there is a simpler framework to determine the OP,
which makes no assumption about the condensate energy and only on
the kinetic energy. We call it {\it the ground state condition}. It
was firstly noticed by A. A. Abrikosov, in his fundamental treatment
of the Ginzburg-Landau theory \cite{abrikosov57}. He obtained the OP
using this condition and found that it also provides an exact
solution of Amp\`ere's law, because it directly relates the
supercurrent to the superconducting density. The ground state
condition does determine the OP and its most relevant aspects, such
as the vortex lattice and the magnetization of the superconducting
state. Thanks to the ground state condition Abrikosov found that the
magnetization is proportional to the spatial average of the OP,
$\langle|\psi|^2\rangle$, where $\psi$ is the one-component OP,
without invoking the Ginzburg-Landau theory. The ground state
condition is expressed by two first order differential equations
that  many years after were rediscovered by E.
Bogomolny~\cite{bogomolny76} in the context of string theory. He
showed that they solve exactly the Ginzburg-Landau theory for a
special coupling value, $\kappa=1/\sqrt{2}$. The two equations are
given by,
\begin{eqnarray}
&&D_{+}\psi=0,\label{gscgl1}, \quad \mbox {and},\\
&&h_3=H-\frac{hq}{mc}{\vert \psi \vert}^2.\label{gscgl2}
\end{eqnarray}
A uniaxial symmetry along the applied field direction must be and
here it is set along the $x_3$ axis, such that $\psi(x_1,x_2)$ and
$h_3(x_1,x_2)$ ($h_1=h_2=0$), are determined by these equations
($D_{+} = D_1+iD_2$, $\vec D\equiv \frac{\hbar}{i}\vec \nabla -
\frac{q}{c}\vec A$, $A_1(x_1,x_2)$, $A_2(x_1,x_2)$). An iterative
way to obtain a solution from these nonlinear equations is to
firstly solve Eq.(\ref{gscgl1}) for $\psi$, under the assumption of
a constant applied external field $H$ ($A_1=0$ and $A_2=Hx_1$). The
first equation is just the lowest Landau level condition whose
solution is $\psi = \sum_k c_k \exp{\big [ik x_1 -\frac{qH}{2\hbar
c}\left(x_2+\frac{\hbar ck}{qH}\right)^2\big ]}$. The set of
wavenumbers $k$ and the constants $c_k$ are determined by imposing
periodic conditions to the order parameter and fixing the number of
vortices within the unit cell area~\cite{doria10}. Next one obtains
$h_3$ from Eq.(\ref{gscgl2}) using the previously determined
$|\psi|^2$. The procedure can be recursively repeated until
convergence is achieved. However just the first step is known to
provide an excellent description of the full GL free energy solution
for $\psi$ and $h_3$ in the range $0.5 H_{c2} \leq H \leq H_{c2}$,
as shown by E.H. Brandt~\cite{brandt95}. Based on the fact that they
are independent of the Ginzburg-Landau theory and its applicability
is not restricted to a single value of $\kappa$, we conclude that
the ground state condition lives in a level more fundamental than
that of the free energy expansion. Nevertheless it is not a
replacement to the free energy expansion since it describes the
vortex lattice, but without determining its symmetry, that can only
be known through a minimization procedure of the free energy
expansion.

Behind the ground state condition is the so-called
Weitzenb\"ock-Liechnorowitz identity, that in the present case is
given by the following expression:
\begin{eqnarray}
F_k= \int \frac{d^2x}{A}\frac{1}{2m}\big({\vert D_1
\psi\vert}^2+{\vert D_2 \psi\vert}^2 \big) =\int
\frac{d^2x}{A}\big(\frac{1}{2m}{\vert D_{+} \psi\vert}^2+\frac{\hbar
q}{2 m c} h_3 {\vert \psi \vert}^2\big). \label{wl1}
\end{eqnarray}
This identity provides a twofold description of the kinetic energy
density, $F_k$. The area orthogonal to the field direction is
described by $A$. This twofold formulation of the kinetic energy
leads to a twofold formulation of the supercurrent, which follows
from the linear term in the vector potential of the kinetic energy,
\begin{eqnarray}
F_k = \int \frac{d^2x}{A}\big(-\frac{1}{c}\, \vec J\cdot \vec
A+\cdots \big).\label{jagl}
\end{eqnarray}
The first formulation of the kinetic energy gives that,
\begin{eqnarray}
J_a&=&\frac{q}{2m}\big[\psi^*\big(D_a\psi\big)+\psi\big(D_a\psi\big)^*
\big], \label{curra}
\end{eqnarray}
$a=1$ or $2$, whereas it follows from the second one that,
\begin{eqnarray}
J_1&=&\frac{q}{2m}\big[\psi^*\big(D_{+}\psi\big)+\psi\big(D_{+}\psi\big)^*
\big]+\frac{\hbar q}{2 m}\partial_2{\vert \psi \vert}^2, \\
\mbox{and} \label{currb1}\nonumber \\
J_2&=&\frac{q}{2m\,\imath}\big[\psi^*\big(D_{+}\psi\big)-\psi\big(D_{+}\psi\big)^*
\big]-\frac{\hbar q}{2 m}\partial_1{\vert \psi \vert}^2.
\label{currb2}
\end{eqnarray}
respectively. Once granted the symmetry along the third axis,
Amp\`ere's law is simply given by $\partial_1 h_3=-4\pi J_2/c$ and
$\partial_2 h_3=4\pi J_1/c$. Then it becomes straightforward to
check that the ground state condition solves Amp\`ere´s law.

The ground state condition is also useful to find a reliable, though
approximate, solution of the other Ginzburg-Landau equation:
\begin{eqnarray}
\frac{1}{2m}\big(D_1^2+D_2^2 \big)\psi =
\alpha(T)\psi-\frac{\beta}{2}|\psi|^2\psi,
\end{eqnarray}
where $\alpha(T)=\alpha_0(T_c-T)$ and $\beta>0$. Integration of this
equation, by firstly multiplying it by $\psi^*$, and then using
periodic boundary conditions, gives that,
\begin{eqnarray}
\int \frac{d^2x}{A}\big[\frac{1}{2m}\big({\vert D_1
\psi\vert}^2+{\vert D_2 \psi\vert}^2\big) -\alpha(T){\vert
\psi\vert}^2 +\beta{\vert \psi\vert}^4\big]  = 0.
\end{eqnarray}
The ground state condition turns this integrated equation into an
algebraic equation, by use of the kinetic energy expression,
Eq.(\ref{wl1}), and the ground state equations, Eqs.(\ref{gscgl1})
and (\ref{gscgl2}):
\begin{eqnarray}
\frac{\hbar q}{2 m c}\big[ H - H_{c2}(T) \big ] \langle
{\vert\psi\vert}^2 \rangle + \beta\big(1-\frac{1}{2\kappa^2}
\big)\langle{\vert\psi\vert}^4\rangle = 0. \quad \label{glhc2}
\end{eqnarray}
The average value means $\langle \cdots \rangle \equiv \int d^2x
\{\cdots \}/A$, the upper critical field is
$H_{c2}(T)=(2mc/q\hbar)\alpha(T)$ and
$\kappa=\sqrt{\beta/2\pi}(mc/\hbar q)$. From this equation one can
easily conclude that if $\kappa > 1/\sqrt{2}$ a non-zero $\psi$
solution is only possible if $H<H_{c2}(T)$. Thus we find that it is
possible to determine the upper critical line ($H_{c2}(T)$) without
explicit calculating the OP, instead, just assuming that the OP
satisfies the ground state equations. Hence the ground state
condition solves exactly Amp\`ere's law and also gives relevant
information about the OP solution of the other Ginzburg-Landau
equation.

We stress the intimate connection between the ground state condition
and the kinetic and field energies, but not to the condensate
energy. This makes the ground state approach independent of the
critical temperature, whose value is determined by the condensate
energy, not present in our considerations. Therefore the ground
state condition applies both below and above the critical
temperature value as solely reflects properties of the kinetic and
field energies.  The ground state condition can also be derived from
the Virial theorem of superconductivity~\cite{doria89,alfredo12}.

Interestingly the ground state equations also appear in the
microscopic approach. Long ago the magnetic field distribution of
pure type-I superconductors with small magnetization was derived
from the non-local version of Gorkov's theory~\cite{brandt74} and
the result is that the local field is equal to the applied field $H$
added to the average gap square~\cite{brandt74,delrieu72}, as
described by Eq.(\ref{gscgl2}).

For all the above reasons we find relevant the derivation of the
ground state condition for the new superconductors. As shown in this
paper this derivation follows from a principle of local rotational
invariance, which is a local gauge symmetry.

\section{The ground state condition for the layered superconductors}

In the previous section we have described the ground state condition
for the traditional superconductors and shown that these equations
stem from the kinetic and field energies. The kinetic energy
describes how the OP is coupled to the local vector potential
through {\it minimal coupling} in order to be gauge invariant. The
steps followed before also apply here to determine the two-component
OP, $\Psi$, and the local magnetic field, $h^k$, in presence of a
spin-charged background. These steps lead to the following
equations:
\begin{eqnarray}
&&\sigma^j(x)D_j\Psi =0 \label{sw1},\quad  \mbox{and}, \label{sw2a}\\
&&h^k = H^k -
\frac{hq}{mc}\Psi^{\dagger}\sigma^k(x)\Psi.\label{sw2b}
\end{eqnarray}
Notice that these equations also do not include the critical
temperature. The spin and charge degrees of freedom of the
background enter the equations through the local Pauli matrices,
$\sigma_j(x)$, and the {\it spin connection} field in the covariant
derivative $D_j$. This covariant derivative is different from the
previous one because besides the local magnetic potential it also
contains a new gauge field to describe the interaction with the
spin-charged background. The study of the local spin and covariant
derivative operators is done in the following sections.

Eqs.(\ref{sw2a}) and (\ref{sw2b}) reflect profound conceptual
changes on the macroscopic approach of superconductivity as compared
to Eqs.(\ref{gscgl1}) and (\ref{gscgl2}). They describe the
interaction between the OP and the local magnetic field over a
spin-charged background defined on a curved space with torsion. Long
ago \'Elie Cartan~\cite{cartan22} developed the formalism of a
curved space with torsion and here we show that it provides the
appropriate venue to include spin correlations and charged
dislocations.

A few years before the discovery of spin by Uhlenbeck and Goudsmith,
Cartan introduced the concept of torsion in general relativity as an
intrinsic angular momentum of
matter~\cite{kibble61,sciama62,hehl76}, whose importance would be in
the same footing as mass. We find that superconductors, which
coexist with a spin-charged background, are the true foreground of
Cartan's geometrical theory. \'Elie Cartan´s geometrical
formalism~\cite{cartan22} was inspired on an analogy with mechanics
of elastic media, like a collection of small granules that are
translated and rotated with respect to each other~\cite{hehl07}. We
stress a fundamental conceptual difference between the use of
Cartan's geometrical formalism to Superconductivity and to General
Relativity. In the latter the curvature of space is caused by mass
and its torsion by spin, whereas for the former only spin
correlations curve the space, as felt by the superconducting
carriers. As we shall see here that torsion in our system is caused
by charged dislocations. In fact \'Elie Cartan's geometrical
formalism has been applied to solids before. A crystal populated by
sufficiently many dislocations can be described by a continuum field
theory~\cite{hehl07}, which by its turn is formulated as a gauge
theory of dislocations~\cite{kadic83,edelen88,katanaev05}. \'Elie
Cartan's formalism~\cite{katanaev92} expresses this gauge theory as
three dimensional theory of gravity. A crystal with dislocations and
disclinations in Euclidean space can be expressed in curved space
with torsion as described by H. Kleinert~\cite{kleinert89}, via a
singular coordinate transformation. This description of defects has
no local invariance under rotations and therefore is not useful for
the present purposes.

Recently dislocations in graphene have been treated as the torsion
field using this formalism~\cite{juan10}. All previous applications
of the so-called Eistein-Cartan geometry to solids have been done in
the context of crystallographic defects, whereas here we apply this
formalism to describe a  spin-charged background felt by the
superconducting state. A feature of Cartan's geometrical formalism
is to turn a curved space theory, like
gravity~\cite{helal09,helal10a,helal10b}, into a Yang-Mills
theory~\cite{kibble61,sciama62,ivanenko83,kazmierczak09}, which is
known to describe the fundamental internal symmetries of particle
Physics. Nevertheless the internal gauge symmetry of \'Elie Cartan´s
approach to gravity is the group of local space-time symmetry.

Notice that Eq.(\ref{sw2a}) is the Euclidean version of the
three-dimensional Dirac equation, with time replaced by one of the
spatial dimensions~\cite{hasan10}, treated differently from the
other two ones. In the Dirac equation the search for eigenstates
breaks the space-time invariance when time is treated differently
from the spatial coordinates as it enters the solution through the
exponential $\exp{(i E t/\hbar)}$, where $E$ is the energy
eigenvalue to be determined. Similarly, the present Euclidean
treatment takes the third coordinate distinctively from the other
two, as it enters the solution through the exponential
$\exp{(-q_3\vert x_3 \vert)}$, where $q_3$ is an eigenvalue to be
determine through the Euclidean Dirac equation turned into an
eigenstate equation. According to this view the OP behavior
perpendicular to the layer is locked to the behavior along the
layer, an important feature for the treatment of multiple layers.
This is done through these equations by linear superposition of the
individual layers, thanks to the linearity of Eq.(\ref{sw2a}) in
$\Psi$. The interaction between layers will carry this
interdependency between in and out of the layer behavior found here,
that may become a fundamental feature to enhance superconductivity,
as stressed by some
authors~\cite{innocenti10,innocenti11a,innocenti11b}. The presence
of the background and of the applied field perpendicular to the
layers yields non-zero solutions for the OP in Eq.(\ref{sw2a}), and
therefore helps to stabilize superconductivity. By its turn
Eq.(\ref{sw2b}) shows that the non-zero OP induces a local magnetic
moment along the layers, given by components
$\Psi^{\dagger}\sigma_1(x)\Psi$ and $\Psi^{\dagger}\sigma_2(x)\Psi$.
However the total magnetization summed over all layers must average
to zero~\cite{doria10}.

We point out to the similarities of the present state with the FFLO
state, which exhibits inhomogeneous superconducting phases
intercalating spatially oscillating OP and spin
polarization~\cite{larkin64,fulde64,matsuda07}. This is similar to
the present OP corrugated by spin correlations and charged
dislocations described by the background. We notice that the ground
state condition is a three-dimensional version of the well-known
Seiberg-Witten equations~\cite{witten94,seiberg94,jost08},
originally written for a smooth compact four dimensional manifold.

\begin{figure}[h]
\centering
\includegraphics[width=1.00\linewidth]{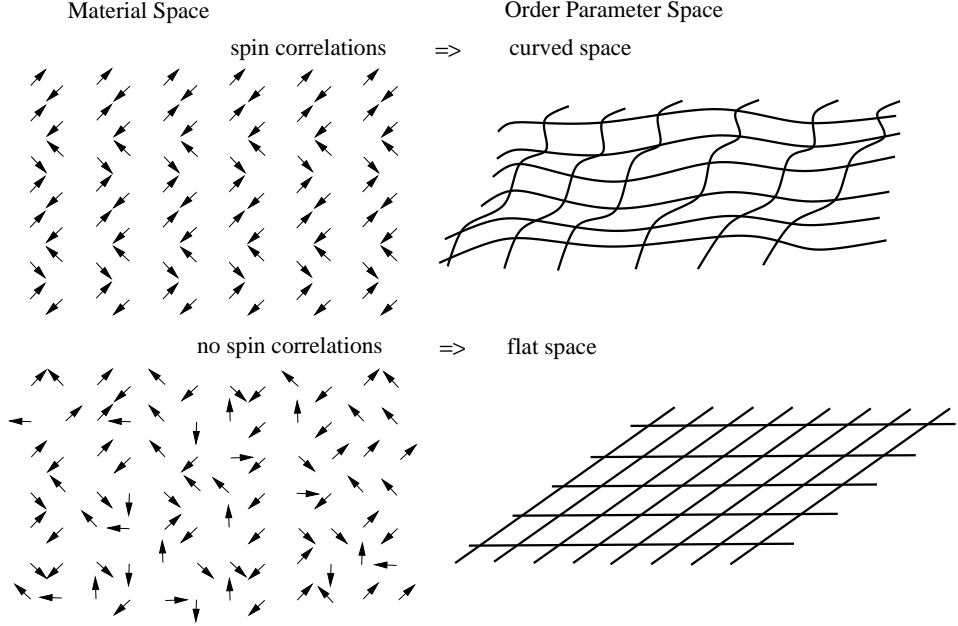}
\caption{This is a pictorial view of the differences between the
material and the order parameter spaces. The observer's view is in
material space where individual charges and spins of static and
mobile carriers interact. The order parameter (OP) space is that
seen by the superconducting condensate which senses the space curved
in regions of intense spin correlations. In regions of uncorrelated
spins the space is flat. This follows from the spin-frame, which in
uncorrelated regions is a pure rotation, $e^i_b(x)= u^i_b(x)$ and
the metric, according to Eq.(\ref{gee}), becomes $g^{i j}=\delta^{i
j}$, to describe a flat space. }\label{fig2}
\end{figure}

\begin{figure}[h]
\centering
\includegraphics[width=1.00\linewidth]{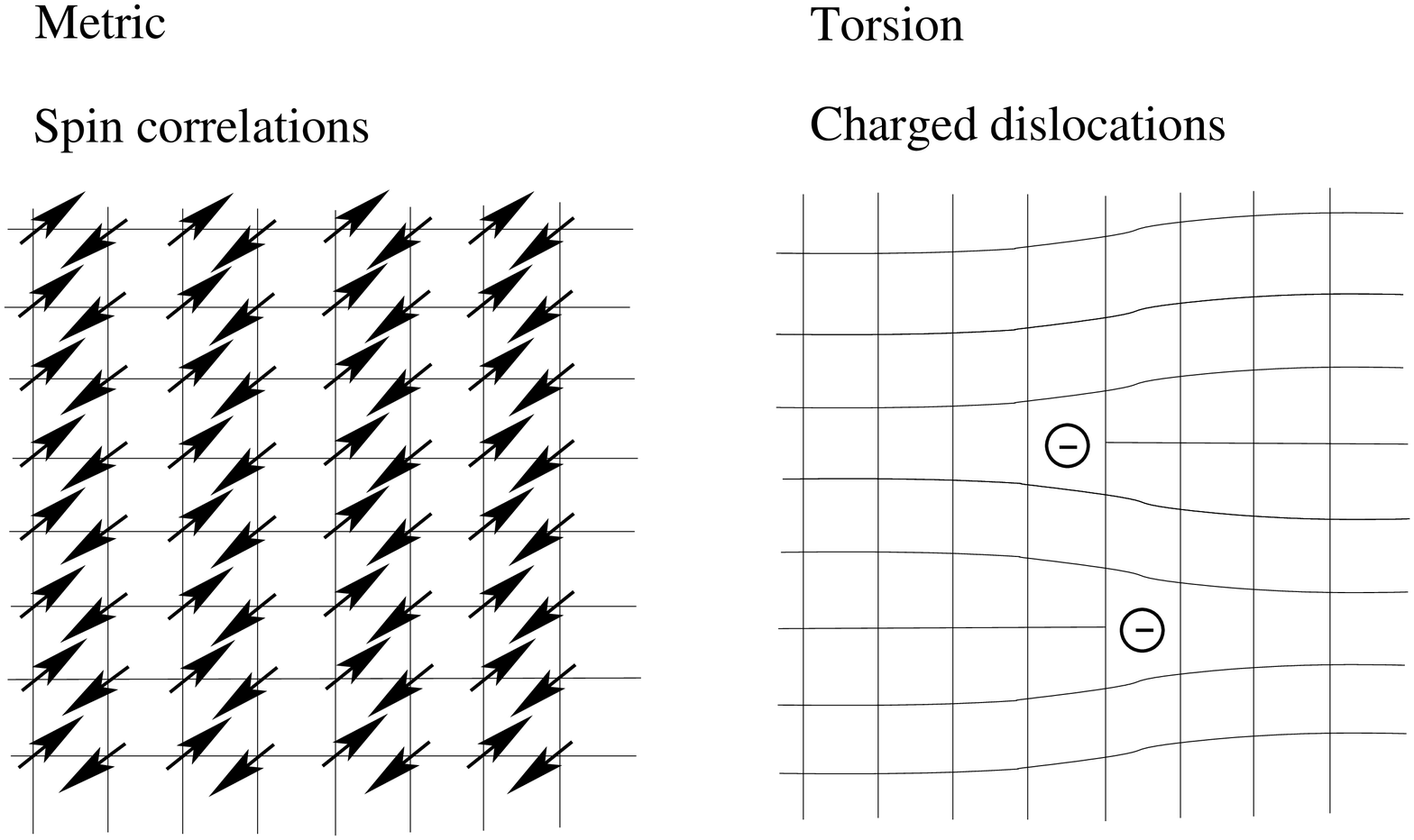}
\caption{Spin correlations and charge dislocations are the the basic
quantities that interact with the order parameter (OP). They are
mathematically described by the metric and the torsion. The
spin-frame, $e^i_a(x)$ is one of the building blocks of Cartan's
geometrical approach, from where the metric follows,
$g^{ij}=e^i_ae^j_a$. The torsion is the other building block and
corresponds to the antisymmetric part of the affine connection,
${\Gamma_{k\,l}}^{i}\varepsilon^{k l m}$.}\label{fig3}
\end{figure}

\section{The local momentum and spin operators \label{lmso}}
The derivation of the ground state equations Eqs.(\ref{sw2a}) and
(\ref{sw2b}) is set over a kinetic energy expression which has a
twofold formulation obtained from a Weitzenb\"ock-Liechnorowitz
identity. Therefore we also follow the same pathway for the
two-component OP in presence of the spin-charged background. The
kinetic energy displays the remarkable property of a {\it
non-abelian gauge symmetry under the group of spatial rotations}.
Therefore all results obtained here follow from this principle of
local rotational invariance. The mathematical details related to the
derivation of the corresponding Weitzenb\"ock-Liechnorowitz
identity~\cite{jost08,cdoria05} will be seen elsewhere. This
identity strongly relies on the commutativity between the local spin
($\sigma^j(x)$) and the local momentum operators ($D_i$). These
operators carry information about the spin-charged background, and
is their locality that render the theory locally invariant under
rotations.

Recently a curved space has been used to treat the effect of
nematicity in superconductors~\cite{barci11}, where "nematic order"
stands for an spontaneously broken rotation symmetry of a lattice
system. A suggestive argument to show the relevance of curved spaces
in strongly correlated systems stems from Fermi liquid theory. This
theory, proposed by Landau in 1956, describes weakly correlated
fermions. It was successfully applied to many systems, such as
Liquid He-3, electrons in a normal metal and protons and neutrons in
an atomic nucleus. A basic concept behind Fermi liquid theory is
that of a quasiparticle, which is a particle dressed by its
interaction with the other particles, resulting into a state with
characteristics of a free particle, namely, a definite momentum and
spin. Both quasiparticles and particles are characterized by
momentum, $\vec P$, and spin, $\vec \sigma$, and this, obviously,
relies on the fact that these two quantities can be simultaneously
observed because they are commuting operators:
\begin{eqnarray}
[P_a,\sigma_b]=0. \label{commps0}
\end{eqnarray}
Assume, for the sake of the argument, that the inhomogeneous spin
state of a strongly correlated system turns the spin into a local
operator described by $\vec \sigma(x)$. The full meaning of this
locality is discussed further in the text. The point to make here is
that this local dependency breaks the above relation,
$[P_i,\sigma^j(x)]\neq 0$, and consequently spoils the concept of a
quasiparticle characterized by momentum and spin. The remarkable
fact is that it is possible to heal this relation and define a new
momentum operator, hereafter called $\nabla_i$ able to commute with
the local spin operator.
\begin{eqnarray}
[\nabla_i,\sigma^j(x)]=0.\label{commps}
\end{eqnarray}
This commutativity is made possible because space has acquired
curvature and torsion by the presence of the spin correlations and
charged dislocations, respectively. The  above relation is the heart
of the present geometrical approach because the momentum operator
$\nabla_i$ is the so-called {\it covariant derivative} introduced by
Fock and Ivanenko in 1929 in the context of General
Relativity~\cite{fockiva29} to deal with Dirac spinors. A similar
but equivalent condition to the commutativity of Eq.(\ref{commps})
was independently introduced by \'Elie Cartan~\cite{cartan22}, whose
formulation of General Relativity is in terms of the so called
"co-frame", a matrix that lives in a more fundamental level then the
metric itself. This matrix plays a fundamental role in the present
treatment and for this reason we call it "spin-frame", since it
contains information about the spin correlations present in the
system.

To construct the local commuting momentum and spin operators we
firstly digress from superconductivity to discuss the OP of a spin
glass, which was shown long ago by many authors to contain more than
one spin field~\cite{edwards75,rivier82,orland83,parisi83,beath07}.
Correlations and not individual spins are important for the spin
glass and here we take the same point of view. Thus we define the
following quantum correlation between a spin in position $x$ and a
reference spin:
\begin{equation}
e^i_a(x)=\langle \, 0 \, \vert \frac{1}{2}\big \{
\sigma^i(x),\sigma_a \big \}\vert \, 0 \, \rangle. \label{equant}
\end{equation}
For the sake of the argument we assume knowledge of the
microscopically obtained spin ground state, $\vert \, 0 \rangle $.
The relevant fact is that the above equation provides a definition
of the spin-frame, $e^i_a(x)$, the building block of Cartan's
geometrical approach. Notice that a point $x$ in $e^i_a(x)$ really
describes an average value over microscopic spins and therefore
refers to a region of degrees of freedom, instead of a single point
in space. The microscopic state $\vert \, 0 \rangle$ is fundamental
to help us understand how correlations enter the problem, but its
derivation is beyond the scope of the present study. For this reason
we simply introduce the following definition of the spin-frame, a
transformation of the Pauli matrices into a new local set,
\begin{equation}
\sigma^i(x)=e^i_b(x)\sigma^b. \label{frame}
\end{equation}
This transformation is consistent with the formulation through the
quantum correlation given in Eq.(\ref{equant}). Nevertheless it must
be kept in mind that this new local set of Pauli matrices does not
represent a single spin as the original set does, and truly carries
information about local spin correlations that must capture the
relevant physical features ranging from a glassy state with
frustrated spins to a state of independent uncorrelated spins.
Hereafter we assume knowledge of the local 3 by 3 spin-frame matrix
$e^i_a(x)$ everywhere. While for a highly correlated spin system the
spin-frame is a full matrix, for the uncorrelated case it becomes an
orthogonal matrix, $e^i_a(x)=u^i_a(x)$. This is because if all spins
are independent rotations of the reference spin, then $\sigma^i(x)=
u^i_b(x)\sigma^b$, where $u^i_b(x)$ represents a rotation in space,
$u^i_a(x)u^i_a(x)=\delta^{i j}$.

Throughout this paper we employ the Einstein notation, that repeated
indices mean a summation. A simple and naive way to describe an
intense correlated spin background, is to assume a scaling function,
$\lambda^{-1}(x)$, $e^i_a(x)=\lambda^{-1}(x) u^i_a(x)$, but one must
bear in mind that this intensity is not of an individual spin,
because it collects correlations contained in the microscopic wave
function $\vert \, 0 \rangle $. (The limit $\lambda\rightarrow 0$
corresponds to an intense spin correlation).

The construction of the local momentum and spin operators
inescapably leads to two fundamental distinct views of space,
hereafter called M (material) and  OP (order parameter) spaces. We
have seen that remarkably the OP space, like M space, also has
commuting momentum and spin operators, and so, quasiparticles over
the correlated spin-charged background can be labeled by them. The
price to pay to have such operators is that the OP space becomes a
curved space. The M space is that of the atoms, periodically
arranged, where spins and charges hop from on site to the other.
Such view is depicted in  Fig.(\ref{fig2}). For instance, the
Hubbard model provides a view of the Mott insulator in M space. The
OP space is that of the macroscopic superconducting state, which is
intrinsically delocalized, and does not feel the individual sites,
but the average correlation defined by Eqs.(\ref{equant}) (quantum
view) and (\ref{frame}) (classical view). We introduce a notation to
distinguish indices associated to M and OP spaces, although both are
three-dimensional: $a,b,c,d=1,2,3$ and $i,k,j,l=1,2,3$ correspond to
M and OP spaces, respectively. Notice that an inverse transformation
can be defined by $\sigma_j(x)=e_j^b(x)\sigma_b$, and because
$\sigma^j\sigma_j=\sigma^b\sigma_b$, we have the general properties
$e^i_a(x)e_j^a(x)=\delta^i_j$, $e_j^b(x)e^j_a(x)=\delta^b_a$, where
$\delta$ always refers to the Kronecker delta (identity matrix).
There is no need to distinguish between upper and lower indices  in
M space ($\sigma^a=\sigma_a$) and throughout this paper our choice
between them is purely done on aesthetical grounds. But in OP space
the situation is different, one must distinguish between them,
($\sigma^i\neq \sigma_i$), as upper and lower indices describe the
transformed set and its inverse, respectively. We shall also refer
to the upper (lower) indices as contravariant (covariant) indices.

For later purposes we briefly review well known features of the
momentum and spin operators of an independent particle. (1a) The
position space representation of the momentum operator is
$P_a\equiv(\hbar/\imath)\partial_a$. The spin operator is described
by the Pauli matrices, $\sigma_a$, whose anti-commutator and the
commutator relations are given by $\{\sigma_a,
\sigma_b\}=2\delta_{a\,b}$, and $\left [\sigma_a, \sigma_b
\right]=2i\varepsilon_{a\,b\,c}, \sigma_c$, respectively. (2a) The
momentum and the spin operator components commute according to
Eq.(\ref{commps0}). (3a) In real space the momentum operator
satisfies the product (Leibniz) rule, $P_a
(\phi_{index}\chi_{index\prime})=(P_a
\phi_{index})\chi_{index\prime}+\phi_{index} (P_a
\chi_{index\prime})$. (4a) The momentum operator components are
commutative, $[P_a,P_b]=0$. The tensors $\phi_{index}$ and
$\chi_{index\prime}$ contain sets of indices, $index$ and
$index\prime$, with an arbitrary combinations of spatial
($a,b,c,d=1,2,3$) and spinorial ($\alpha,\beta=1,2$) indices.
Consistency with (1a) and (2a) implies that  the action of the
momentum operator on some special tensors gives zero, namely, the
Kronecker delta, $P_a\, \delta_{b\,c}=0$, and the totally
anti-symmetric tensor, $P_a\,\epsilon^{b\,c\,d}=0$. This last tensor
takes values $\epsilon^{b\,c\,d}=1,-1,0$, for cyclic, anti-cyclic
and repeated indices, respectively.

Next we explore the consequences of the local spin operator defined
in OP space, due to Eq.(\ref{frame}). The anti-commutation and the
commutation of the Pauli matrices become,
\begin{eqnarray}
\{\sigma^i(x), \sigma^j(x)\}=2g^{ij}(x), \label{acomm} \\
g^{ij}(x)=e^i_a(x)e^j_a(x), \label{gee}\\
\left [\sigma^i(x), \sigma^j(x) \right] = 2i\varepsilon^{i\,j\,k}(x) \sigma_k(x),\label{comm}\\
\varepsilon^{i\,j\,k}(x)=e^{-1}\epsilon^{i\,j\,k}. \label{eps}
\end{eqnarray}
The determinant of the spin-frame satisfies the condition
\begin{eqnarray}
e^i_a e^j_b e^k_c
\epsilon^{a\,b\,c}=e^{-1}\epsilon^{i\,j\,k}.\label{det}
\end{eqnarray}
Notice that Eq.(\ref{comm}) contains both the transformation and its
inverse, the left side of it the transformed spins, whereas its
right the inverse of this transformation. The determinant of $e^i_a$
is denoted by $e^{-1}$, and is given by Eq.(\ref{det}), which
contains the totally anti-symmetric tensors $\epsilon^{a\,b\,c}$ and
$\epsilon^{i\,j\,k}$, taking values 1,-1 and 0. However
$\epsilon^{a\,b\,c}$ is a tensor in M space but $\epsilon^{i\,j\,k}$
is not a tensor in OP space. The antisymmetric tensor in OP,
$\varepsilon^{i\,j\,k}$, is defined in Eq.(\ref{eps}). Thus there
are the two distinct notations, $\epsilon$ and $\varepsilon$, for
the antisymmetric tensors in M and OP space, respectively.

Next we summarize the properties of local momentum and spin
operators in OP space. (1b) There is a momentum operator,
$\nabla_i$, also called the covariant derivative. The spin operator,
$\sigma^i(x)$, has anti-commutator and commutator relations given by
Eqs.(\ref{acomm}) and (\ref{comm}), respectively. (2b) The momentum
and the spin operator components commute according to
Eq.(\ref{commps}). (3b) The momentum operator satisfies the product
(Leibniz) rule, $\nabla_i (\phi_{index}\chi_{index\prime})=(\nabla_i
\phi_{index})\chi_{index\prime}+\phi_{index} (\nabla_i
\chi_{index\prime})$. (4b) The momentum operator components do not
commute, $[\nabla_i,\nabla_j] \neq 0$ because of the spin-charged
background. The tensors $\phi_{index}$ and $\chi_{index\prime}$
contain sets of indices, $index$ and $index\prime$, with an
arbitrary combinations of M space ($a,b,c,d=1,2,3$), OP space
($i,j,k,l=1,2,3$), and spinorial ($\alpha,\beta=1,2$) indices.

Similarly to the free independent particle case, the covariant
derivative applied to the special tensors must also vanish by
consistency or by use of its explicit form~\cite{alfredo12}. The
covariant derivative commutes with both M and OP space spin
operators, namely $ [\nabla_i,\sigma_a]=0$ and Eq.(\ref{commps})
holds, and from this one obtains that,
\begin{eqnarray}
&& \nabla_i \, \sigma_a = 0 \quad  \mbox{and} \\
&& \nabla_i \, \sigma^j(x)=0\label{dersig},
\end{eqnarray}
this last equation being another way to express the Fock-Ivanenko
condition~\cite{fockiva29} given in Eq.(\ref{commps}). From the
anticommutator and the commutator relations, given by
Eqs.(\ref{acomm}) and (\ref{comm}), respectively, it follows that
the tensors defined by Eqs.(\ref{gee}) and (\ref{eps}) satisfy,
\begin{eqnarray}
&&\nabla_i \, g^{j\,k}(x)=0 \quad \mbox{and}  \label{metricity}\\
&&\nabla_i \, \varepsilon^{j\,k\,l}(x)=0. \label{dereps}
\end{eqnarray}
The Kronecker delta in M and OP space vanish under the covariant
derivative, $\nabla_k \big(\delta^j_m \big)=0$, $\nabla_k
\big(\delta^{a b} \big)=0$, and also does the totally antisymmetric
symbol in M space, $\nabla_k \big(\epsilon^{a b c}\big)=0$. The lack
of commutativity between momentum components introduces novel
physical aspects to this theory brought by the spin-charged
background. We recall that in M space this commutativity is also
lost by the presence of a magnetic field. According to
Eq.(\ref{gee}), and the Leibniz rule, the covariant derivative
applied to the spin-frame must vanish,
\begin{eqnarray}
\nabla_k \, e^{i}_a(x)=0.
\end{eqnarray}
The vanishing of the covariant derivative applied to all special
tensor is a straightforward consequence of the above condition,
which is then the most important of all relations. This is the
condition introduced by Cartan which is in an equal footing to
Eq.(\ref{commps}). Notice that all the above relations follow from
consistency without making use of the explicit form of this
covariant derivative, which is only done in the next section. In its
explicit form the above equation is given by,
\begin{eqnarray}
\nabla_k \, e^i_a =  \frac{\hbar}{\imath}\big [\partial_k  e^i_a-g
\omega_{k a b}\,e^i_b+{\Gamma_{k\,m}}^{i} e^m_a \big ], \label{de}
\end{eqnarray}
A detailed discussion of the above expression is carried
elsewhere~\cite{alfredo12}. Notice that the covariant derivative
contains two new sets of fields: (1) {\it the spin connection},
which carries OP and M space indices, $\omega_{iab}$, and is
antisymmetric in $ab$ ($\omega_{iba}= -\omega_{iab}$); (2) {\it the
affine connection} with just OP indices, ${\Gamma_{i\,m}}^{k}$.

The metric tensor $g^{ij}$ that naturally arises in Eq.(\ref{gee})
describes the infinitesimal distance between nearest points in OP
space separated by $dx_i$. To see this just consider the internal
product $\sigma^i dx_i=\sigma^a dx_a$, and take the anticommutator
$\{\sigma^i dx_i,\sigma^j dx_j\}/2=g^{ij}dx_idx_j=\{\sigma^a
dx_a,\sigma^b dx_b\}/2=dx_adx_a$ since in M space the metric is the
Kronecker delta. Therefore we reach the conclusion that OP is curved
and the concept of distance must be defined as
$ds^2=g^{ij}\,dx_i\,dx_j$. We stress that uncorrelated spins do not
curve the OP space. We have argued before that if all spins are
independent rotations of the reference spin, then the spin-frame is
a pure rotation, $e^i_b(x)= u^i_b(x)$ and this renders the metric,
according to Eq.(\ref{gee}), trivial, $g^{i j}=\delta^{i j}$ since
$u^i_a(x)u^j_a(x)=\delta^{i j}$. Two nearest points in OP and M
spaces, are related by $dx^i=e^i_a(x)\;dx^a$, or equivalently,
$e^i_a(x)=\partial x^i/\partial x^a$. From this it follows that
under a general coordinate transformation,
\begin{eqnarray}
{e^{\prime \, i}}_a(x^{\prime})= \frac{\partial x^{\prime
\,i}}{\partial x^a}=\frac{\partial x^{\prime\,i}}{\partial
x^j}\frac{\partial x^j}{\partial x^a}\equiv \Lambda^i_j e^j_a(x).
\end{eqnarray}
Apply the determinant to both sides and consider that $e^{-1}\equiv
det (e^i_a)$, to obtain that $e^{\prime \, -1}=det(\frac{\partial
x^{\prime\,i}}{\partial x^j})e^{-1}$. This means that a volume in OP
space must be corrected according to $e^{\prime} d^3x^{\prime}= e
d^3x$. The determinant, $1/e$, is a measure of the intensity of the
spin correlation $e^i_a(x)$, and shows that regions of high spin
correlation ($e \rightarrow 0$) have less effective volume to
integrate in OP space than those of uncorrelated spins ($e
\rightarrow 1$). Thus spin and superconductivity are indeed
competing effects in this formulation too.

\section{Gauge symmetry of local rotations\label{gslri}}

The derivation of the local momentum and spin operators set over the
spin-charged background is a key element to obtain the kinetic
energy of the condensate~\cite{alfredo12}.  The symbol $D_j$ used to
to express the covariant derivative applied to $\Psi$, as shown in
Eq.(\ref{gscgl1}), is given by,
\begin{eqnarray}
D_j\Psi \equiv \big (\frac{\hbar}{i}\partial_j-\frac{\hbar
g}{2}\omega_{j a b}\Sigma^{a\,b} - \frac{q}{c}A_{j} \big )\Psi.
\label{NPsi}
\end{eqnarray}
Notice the presence of a new coupling constant $g$ to describe the
interaction with the spin-charged background.  The hermitian
matrices $\Sigma_{a\,b}$ are the generators of the rotation group
and satisfy the commutation rule,
$[\Sigma_{a\,b},\Sigma_{c\,d}]=\imath \delta_{a\,c}
\Sigma_{b\,d}+\imath \delta_{b\,d} \Sigma_{a\,c} - \imath
\delta_{a\,d} \Sigma_{b\,c}-\imath \delta_{b\,c} \Sigma_{a\,b}$, but
for the two-component OP they become Pauli matrices,
$\Sigma_{a\,b}=-\imath
[\sigma_a,\sigma_b]/2=\epsilon_{a\,b\,c}\sigma^{c}$. Therefore
besides the spin-frame $e^a_i(x)$, the interaction of the OP with
the background also demands the spin connection $\omega_{j a b}(x)$.
A scalar OP, such as in case of the standard Ginzburg-Landau theory,
can not minimally couple to the spin-charged background, because the
covariant derivative is simply given by
$\nabla_i\psi=\big(\hbar/\imath\big)\partial_i\psi$. Therefore the
superconducting OP must have at least two components to develop
minimal coupling to a spin-charged background. Thus minimal coupling
introduces in the kinetic energy a SU$_L$(2)$\otimes$U$_L$(1)
symmetry, where $L$ stands for local. Under a local rotation
$U(x)=\exp(\imath\theta_{a\,b}\Sigma^{a\,b})$ the OP transforms as
$\Psi^\prime=U\Psi$ and invariance under local rotation,
$\nabla_j(\Psi^\prime)=U\,\nabla^\prime_j\Psi$, sets the way the
spin connection must transform,
${\omega^\prime}_j^{a\,b}\Sigma_{a\,b} = U^{-1}
\omega_j^{a\,b}\Sigma_{a\,b}U - \big(2/\imath g \big)U^{-1}
\partial_j U$. The spin connection plays the role of the vector potential
in the non-abelian gauge (Yang-Mills)
theory~\cite{hehl76,ivanenko83}. Under a U$_L$(1) rotation,
$\Psi^\prime=\exp(\imath\theta)\Psi$, the vector potential
transforms, as below, to have the gauge invariance of
electromagnetism: $ {A^\prime}_j = A_j - \big(c\hbar/q \big)
\partial_j \theta$. The commutator becomes,
\begin{eqnarray}
&&\big[ D_i,D_j \big ] = -\frac{\hbar q}{\imath c} F_{i\, j}-\frac{\hbar^2 g}{2 \imath} R_{i\, j}^{a \, b} \Sigma_{a\,b},\\
&&F_{i\,j} = \partial_i A_j - \partial_j A_i,  \label{fij} \\
&&R^{a\,b}_{i\,j}(\omega) =  \partial_i \omega^{a\,b}_j -
\partial_j \omega^{a\,b}_i - g \big ( \omega_i^{a\,c} \omega_j^{b\,c}
-\omega_i^{b\,c} \omega_j^{a\,c}\big ).\label{riemanomega}
\end{eqnarray}
The spin-charged background modifies the commutativity of the
momentum components and besides the electromagnetic field
$F_{i\,j}$, there is also the Riemannian curvature tensor
$R^{a\,b}_{i\,j}(\omega)$. Using Eq.(\ref{de}) the second derivative
of the spin frame becomes
$\big(\partial_l\,\partial_k-\partial_k\,\partial_l\big) e^i_a= g
R^{a\,b}_{l\,k}(\omega)e^i_b-{R_{l\,k\,j}}^{i}(\Gamma)e^j_a $. We
impose the condition that the spin-frame has a smooth spatial
behavior, such that
$\big(\partial_l\,\partial_k-\partial_k\,\partial_l\big) e^i_a=0$,
to obtain that, $ g R^{a\,b}_{l\,k}(\omega)e^i_b
e^j_a={R_{l\,k}}^{j\,i}(\Gamma)$. This means that the Riemannian
curvature tensor, already expressed in terms of the spin connection
in Eq.(\ref{riemanomega}), can also be written solely in terms of
the affine connection: ${{R_{l\,k}}_j}^{\,i}(\Gamma)= \partial_l
{\Gamma_{k\,j}}^i -\partial_k {\Gamma_{l\,j}}^i - \big (
{\Gamma_{k\,m}}^i \; {\Gamma_{l\,j}}^m-{\Gamma_{l\,m}}^i \;
{\Gamma_{m\,j}}^m \big)$. In Kleinert's description of defects, the
rotational symmetry is absent, and so, $\omega^{a\,b}_i$ is zero.
Then the curvature necessarily has singular behavior, and so does
the spin-frame field, since in this case one obtains that
${{R_{l\,k}}_j}^{\,i}(\Gamma) =
e_j^a\big(\partial_l\,\partial_k-\partial_k\,\partial_l\big) e^i_a$.
This shows that the Kleinert's limit is not useful for the purpose
of describing the non-singular spin-charged background treated here.
\section{The SU$_L$(2)$\otimes$ U$_L$(1) invariant ground state
condition\label{groundstate2}}

For the two-component OP the kinetic energy can be expressed in two
different but equivalent ways given below:
\begin{eqnarray}
F_k &=& \frac{1}{2m}\int \frac{d^3x}{V} \,e \big \{ g^{i j}\, \big
(D_i \Psi \big)^{\dagger}(D_j \Psi \big) -\hbar^2 g  R \,\Psi
\label{knt1} ^{\dagger} \Psi  - \nonumber \\
&-&\frac{\hbar}{2}{\Gamma_{l\,m}}^{j}\varepsilon^{l m k} \big [\Psi
^{\dagger}\sigma_k( D_j \Psi \big)+ \big( D_j
\Psi\big)^{\dagger}\sigma_k \Psi\big]\big \},\quad \mbox{and,} \\
F_k &=& \frac{1}{2m}\int \frac{d^3x}{V} \,e \big \{\big (\sigma^i
D_i \Psi \big)^{\dagger}(\sigma^j D_j \Psi \big) +\frac{\hbar q}{c}
 h^k \Psi^{\dagger} \sigma_k\Psi \big \}.
\label{knt2}
\end{eqnarray}
The equality between Eqs.(\ref{knt1}) and (\ref{knt2}) is the
Weitzenb\"ock-Liechnorowitz identity~\cite{jost08}. The derivation
of this identity will be given elsewhere~\cite{alfredo12}. We stress
that such derivation relies on the commutativity between local spin
and momentum operators, this last one being essentially the
covariant derivative. The most natural way to express the kinetic
energy is through Eq.(\ref{knt1}), because of the gradient square
term, $(D_i \Psi \big)^{\dagger}(D_j \Psi \big)$. Notice the
presence of new and important contributions brought by the
spin-charged background, namely, the scalar Riemannian curvature
$R$, and the torsion field ${\Gamma_{l\,m}}^{j}\varepsilon^{l m k}$.
The local magnetic field and the Riemannian scalar curvature are
given by,
\begin{eqnarray}
h^k &\equiv& \frac{1}{2}\varepsilon^{i j k}F_{ij} \quad \mbox{and}\\
R &\equiv& e^i_a e^j_b R^{ab}_{ij},
\end{eqnarray}
respectively. Eq.(\ref{knt1}) has a temperature like term,
$gR\Psi^{\dagger}\Psi$, in the kinetic energy of the condensate,
which embodies our description of inhomogeneities by the present
formalism. It is well-known that in the Ginzburg-Landau theory a
term proportional to $|\Psi|^2$ sustains superconductivity because
it flips sign in the critical temperature, being negative below and
positive above. The Riemannian scalar curvature $R$ also changes
sign from one region to another, depending on the spin-charged
background, leading to a naturally heterogeneous order parameter,
and so, to inhomogeneous superconductivity.

The electromagnetic current follows from the linear term of the
kinetic energy, proportional to the vector potential,
\begin{eqnarray}
F_k = \int \frac{d^3x}{V}\,e\,\big(-\frac{1}{c}J^iA_i+\cdots
\big),\label{jant}
\end{eqnarray}
and consequently there are also two equivalent expressions for it:
\begin{eqnarray}
J^i &=& \frac{q}{2m}g^{i j} \big[ \Psi^{\dagger}\big(D_j\Psi \big)+  \big(D_j\Psi \big)^{\dagger}\Psi  \big ] +\nonumber \\
    &+&  \frac{\hbar q}{2m} {\Gamma_{k\,l}}^{i}\varepsilon^{k l m} \Psi^{\dagger}\sigma_m\Psi,
    \quad \mbox{and}\label{curr1} \\
J^i &=& \frac{q}{2m} \big[ \big( \sigma^j D_j \Psi \big)^{\dagger} \sigma^i \Psi + \Psi^{\dagger} \sigma^i \big( \sigma^j D_j \Psi \big)\big ] - \nonumber \\
 &-& \frac{\hbar q}{2 m}\varepsilon^{i j k} \partial_j\big(
 \Psi^{\dagger}\sigma_k\Psi\big),\label{curr2}
\end{eqnarray}
At this point it becomes clear how the background interacts
electromagnetically with the superconducting OP.  The kinetic energy
and the supercurrent expressions, given by Eq.(\ref{knt1}) and
Eq.(\ref{curr1}), respectively, show that the torsion field,
${\Gamma_{k\,l}}^{i}\varepsilon^{k l m}$, known to describe
dislocations in a crystal~\cite{hehl07,kleinert89,katanaev92},
contributes electromagnetically to the supercurrent according to
Eq.(\ref{curr1}). To obtain Amp\`ere's law firstly take the field
energy,
\begin{eqnarray}
F_f = \int \frac{d^3x}{V}\,e\,\frac{h^k h_k}{8\pi},\label{jant2}
\end{eqnarray}
and then the current contribution to the kinetic energy, to obtain
that,
\begin{eqnarray}
\varepsilon^{i j k}\partial_k h_i=\frac{4\pi}{c}J^i. \label{amp2}
\end{eqnarray}
The contravariant magnetic field $h^k$ defines the {\it magnetic
induction} $B^k$, because the volume integral turns into a line
integral by means of $\varepsilon^{klm}=\epsilon^{klm}/e$. The
magnetic induction is a topological number that counts the number of
vortices, and so, is proportional to an integer vector $\vec N =
(N_1,N_2,N_3)$. For a cubic unit cell with size $V^{1/3}$, with
periodic boundary conditions~\cite{doria89}, one obtains that $ B^k
\equiv \int d^3x\,e\,h^k/V = \int d^3x\,\epsilon^{klm}\partial_l
A_m/V = 2\pi\Phi_0 N^k/V^{2/3}$. Thus we reach the conclusion that
ground state equations solve exactly Amp\`ere's law,
Eq.(\ref{amp2}), thanks to the current, Eq.(\ref{curr2}) because the
first ground state equation, Eq.(\ref{sw2a}) leads to the second
one, Eq.(\ref{sw2b}).

Gauge field theories have been proposed to describe disordered
systems~\cite{dzyaloshinskii78,hertz78,sarkar80,rivier82} because a
disordered system, in its most general form, breaks the
translational invariance in every point of space, but locally
preserves the rotational invariance. Therefore at each point one
defines a new set of axis, and so, there is local rotational
invariance which becomes a non-abelian gauge symmetry with the
rotation group playing the role of the internal
symmetry~\cite{rivier82}. Essentially we have extended this proposal
to describe disordered systems to the superconducting state in
presence of a spin-charged background.


\section{Conclusion\label{con}}

The present approach describes how the order parameter feels an
inhomogeneous background. For the one-component OP, $\psi$, such
inhomogeneous background can be thought as a local distribution of
critical temperatures, $T_c(x)$, that turns the condensate energy,
$\alpha_0(T-T_c(x)){\vert \psi\vert}^2 +\beta{\vert \psi\vert}^4/2$,
local. Some regions will have the temperature above the local
critical temperature, $T>T_c(x)$, whereas others not, $T<T_c(x)$.
The condensate energy can be locally positive or negative,
suggesting a possible coexistence of normal ($\psi=0$) and
superconducting regions ($\psi \ne 0$). However a full response to
the size of these regions requires that we take into account the
kinetic energy, which connects all points into a single state. Thus
one must add the kinetic to condensate energy to find the energetic
cost of interfaces separating the normal to the superconducting
regions. For the present case description of a two-component OP,
$\Psi$, we have obtained in this paper a new mechanism to locally
sustain superconductivity based only on the kinetic energy. This
mechanism relies on the fact that the kinetic energy contains a term
$R\vert \Psi \vert^2$, where the Riemannian spatial curvature $R$ is
induced by spin correlations and charge dislocations. This curvature
can be either positive or negative, and similarly to a local
critical temperature turns -$R\vert \Psi \vert^2$ into a locally
negative or positive term. This term added to the traditional
gradient square term present in the kinetic energy, as seen in
Eq.(\ref{knt1}), sustains a spatially corrugated OP.

In summary we propose here an approach to describe a superconducting
layer through a two-component order parameter interlaced with extra
spin and charge degrees of freedom. The order parameter and the
local magnetic field are determined by first order differential
equations in presence of the spin-charged background. The obtainment
of local momentum and spin operators, constructed over the
spin-charged background, leads to a twofold view of the kinetic
energy because of the Weitzenb\"ock-Liechnorowitz identity. This
also leads to a twofold formulation of the supercurrent, and to the
solution of Amp\`ere's law, yielding the sought first order
differential equations. The local commuting spin and momentum
operators show that the order parameter lives in a curved space with
torsion, as described by the geometrical approach of \'Elie Cartan.
The ground state displays a non-abelian gauge symmetry set by local
invariance under rotations. The ground state equations are meant to
describe the order parameter ranging from the strong spin
correlation regime to the charged dislocation regime.

\section*{Acknowledgments} This work is supported by the brazilian agencies
CNPq, Faperj and Facepe (grant
0589/1.05-08). 

\end{document}